\documentclass[a4paper,12pt]{article}
\usepackage[left=2cm,right=2cm]{geometry}
\usepackage{graphicx}
\usepackage{mathrsfs}
\usepackage{amsmath}
\usepackage[scriptsize,bf]{subfigure}
\newcommand{\sla}{\displaystyle{\not}}

\title{The $N^*(1710)$ as a resonance in the $\pi\pi N$ system}
\author{K. P. Khemchandani\footnote{kanchan@ific.uv.es}, A. Mart\'inez Torres\footnote{amartine@ific.uv.es}, and E. Oset\footnote{oset@ific.uv.es}}
\date{}
\begin{document}
\maketitle
\begin{center}
Departamento de F\'{\i}sica Te\'orica and IFIC,
Centro Mixto Universidad de Valencia-CSIC, Institutos de
Investigaci\'on de Paterna, Aptdo. 22085, 46071 Valencia, Spain
\end{center}
\abstract{We study the $\pi \pi N$ system by solving the Faddeev equations, for which the input two-body 
$t$-matrices are obtained by solving the Bethe-Salpeter equation in the coupled channel formalism.
The potentials for the $\pi \pi$, $\pi N$ sub-systems and their coupled channels are obtained from chiral Lagrangians,
which have been earlier used to study resonances in these systems successfully. 
In this work, we find a resonance in the $\pi\pi N$ system with a mass of $1704 - i 375/2$ MeV and with quantum numbers $I=1/2$, $J^\pi =1/2^+$. We identify this state with the $N^*(1710)$. This peak is found where the energies of the $\pi \pi$ sub-system fall in the region of the $\sigma$ resonance. We do not find evidence for the Roper resonance in our study indicating a more complex structure for this resonance, nor for any state with total isospin $I=3/2$ or $5/2$.}

\section{Introduction}
The excited states of the nucleon have been studied extensively theoretically as well as experimentally. This is evident from the fact that many of these states, especially those in the energy region below 1750 MeV, have been assessed either three or four stars by the particle data group (PDG) \cite{pdg}. Even then, there are some resonances in this low energy region which still need unanimous agreement on their  characteristics or existence, e.g., the $J^\pi = 1/2^+$ resonances in the isospin $1/2$ domain. The $N^*(1440)$ or Roper resonance is a subject of continuous debate and the existence of the $N^*(1710)$ is even questioned. The quark models face difficulties in reproducing both these states \cite{isgur1, isgur2, glozman}. In case of the $N^*(1710)$, some partial wave analyses \cite{arndt1, arndt2} do not find any pole corresponding to it,  while  others claim  a clear manifestation of this resonance \cite{cut, manley, chiang, batinic}. On the other hand, the authors of \cite{ceci} claim an indisputable existence of the $N^*(1710)$ from their study of the $\pi N \rightarrow \eta N$ reaction
in the coupled channel formalism and suggest that the status of this resonance should be improved from three-star to four-star. 

Another controversy about the $N^*(1710)$ started after the finding of a narrow peak in the $\gamma A \rightarrow (K^+ n) X $
reaction at LEPS \cite{nakano}, suggesting the existence of a pentaquark state which some groups associated to a SU(3) antidecuplet to which the $N^*(1710)$ would also belong (see, for example, \cite{diakonov, jaffe}). In order to be compatible with the  $\Theta^+$, the $N^*(1710)$ is required to be narrow. However, the width of this resonance is not known precisely, with the widths listed in \cite{pdg} ranging from $\sim$ 90 - 480 MeV. The authors of \cite{ceci2} re-analyzed the $\pi N \rightarrow K \Lambda$ reaction and found that a narrow width of the $N^*(1710)$ \cite{pdg} was incompatible with the data and proposed the existence of another narrow resonance in this energy region. The partial wave analyses group who do not find a pole for the $N^*(1710)$ suggested to look for other resonance in this energy region as a possible narrow, non-strange partner of the $\theta^+$ \cite{arndt3,arndt4}. The debate on this issue has continued with new analyses which do not find a signal for the $\theta^+$, as a consequence of which, the case for this state has weakend (see \cite{hicks} for a review).

In case of the Roper resonance, which is the lowest excited state of the nucleon and, hence, in the simplest quark model should be expected to be a 3-quark state with a radial excitation of a quark, alternative descriptions, like a 3-quark-gluon structure \cite{q3g}, a quark core dressed by meson clouds \cite{cano}, a dynamically generated resonance from interaction of mesons and a baryon \cite{juelich}, etc., are posed in order to reproduce its properties.

Looking at the characteristics of both these $1/2^+$ resonances in \cite{pdg}, i.e., a large branching ratio for the $\pi \pi N$ decay channel, ($\sim$ 30-40 $\%$ for the $N^*(1440)$ and 40-90 $\%$ for the $N^*(1710)$), it seems that they couple strongly to two meson-one baryon systems. There are many findings which support this idea, e.g., a strong $\sigma N$ coupling to the Roper resonance reported in \cite{juelich, dillig}, an important contribution from the two meson cloud to the masses of the SU(3) antidecuplet members found in \cite{hosaka}, and a good reproduction of the data on the $\Sigma \pi$ distribution in the $\pi^- p \rightarrow K^0 \Sigma \pi$ reaction by taking the $\pi \pi N$ decay channel of the $N^*(1710)$ into account \cite{hyodo}. Hence, a study of the three-body structure of these resonances could shed more light on their properties.

We study the $\pi \pi N$ system by solving Faddeev equations in s-wave using the formalism developed in \cite{mko1, mko2, mko3}. In these works, we studied 
systems made of two mesons and a baryon and those of three mesons by using unitary chiral dynamics to calculate the  two-body amplitudes required for the Faddeev equations. In \cite{mko1, mko2} we have investigated the $\pi \bar{K} N$ system and coupled channels and found a strong coupling of  the lowest lying $1/2^+$ $\Sigma$ and $\Lambda$ resonances of the $PDG$ \cite{pdg} to the three-body decay channels. 
In addition to this, unknown quantum numbers of some of the $S = -1$ resonances were predicted, e.g., $\Sigma(1560)$ has been listed with an unknown spin-parity in \cite{pdg} and our work \cite{mko1} generates it with $J^\pi = 1/2^+$.
Analogously, a three meson system formed by two pseudoscalar mesons, $K$ and $\bar{K}$, and one vector meson, $\phi$, has been studied in \cite{mko3}, revealing the existence of a resonant state in this system at $\sim$ $2150$ MeV when the $K \bar{K}$ invariant mass is close to that of the $f_0(980)$, which can be identified with the $X(2175)$ $1^{--}$ resonance discovered in the process $e^+ e^- \to \phi f_0(980)$ at BABAR \cite{BABAR1,BABAR2} and in the $J/\Psi \to \eta \phi f_0(980)$ reaction by the BES collaboration \cite{BES}.

\section{Formalism} 
In this section we shall discuss different three body interaction diagrams, which contribute to the three-body scattering matrix, starting from the lowest order and going up to the higher order ones. We consider the $\pi \pi N$ system with total charge zero and use $\pi^0 \pi^0 n$, $\pi^+ \pi^- n$, $\pi^- \pi^+ n$, $\pi^0 \pi^- p$ and $\pi^- \pi^0 p$ as coupled channels. Taking advantage of the formalism developed in \cite{mko1, mko2}, where 22 coupled channels were handled simultaneously, we have also used an extended base of states, up to 14, including $\pi K \Sigma$, $\pi K \Lambda$, $\pi \eta N$ channels and we found the results remarkably similar to those obtained using the $\pi \pi N$ channels alone.  Thus we present the formalism for the $\pi \pi N$ states which is considerably simpler. We need the $\pi N$ and $\pi \pi$ $t$-matrices as input, for which we solve coupled channel Bethe-Salpeter equations as explained in the following sub-sections. 

\subsection{Lowest order diagrams}
\begin{figure}
\begin{center}
\includegraphics[scale=0.7]{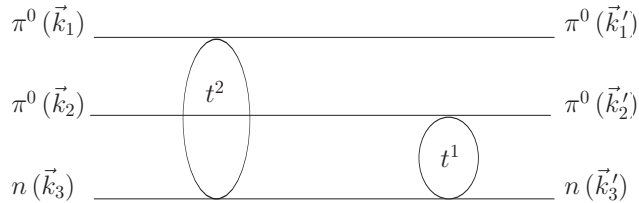}
\caption{\it An example of a simplest possible interaction amongst the three particles, $\pi^0 \pi^0 n$. The labels $\vec{k}_i$ ($\vec{k}_i^\prime$) on the particle lines denote the momenta corresponding to the initial (final) state. The meaning of the blob is shown in Fig. \ref{ls}. }\label{tgt1}
\end{center}
\end{figure}
\begin{figure}
\begin{center}
\includegraphics[scale=0.6]{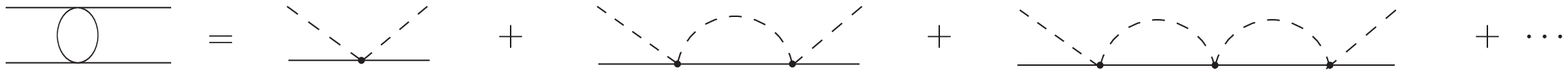}
\caption{\it The blob in the Fig. \ref{tgt1}, which is a $t$-matrix. }\label{ls}
\end{center}
\end{figure}
Our aim is to calculate the three-body scattering matrix which includes all the possible ``connected'' diagrams between the three particles. The simplest possible diagram of this kind is the one which contains two $t$-matrices, for example the one shown in Fig. \ref{tgt1} for the $\pi^0 \pi^0 n$ channel. Following \cite{mko1}, this diagram can be expressed mathematically as ( reading Fig. \ref{tgt1} from right to left as a convention )
\begin{equation}
t^1 g^{12} t^2 = t^1 \frac{M_3}{ E_3 ( \vec{k}_1^\prime + \vec{k}_2 ) }\frac{1}{\sqrt{s} - E_1 (\vec{k}_1^\prime) - E_2(\vec{k}_2  )  - E_3(\vec{k}_1^\prime  + \vec{k}_2) + i \epsilon}\,\, t^2, \label{tgt}
\end{equation}
where the superscript on $t$ denotes the particle which is not interacting in the three-body system. Hence, $t^1$ is the $t$-matrix for the interaction of particles $2$ and $3$ and $t^2$ is that for particles $1$ and $3$. In order to calculate these $\pi^0 n \rightarrow \pi^0 n$ $t$-matrices the Bethe-Salpeter equation,
\begin{equation}
t = v + v\tilde{g} t \label{bs}
\end{equation}
is solved in the coupled channel approach with the potentials obtained from the Lagrangian \cite{weinberg, meissner, ecker, angels}
\begin{equation}
\mathcal{L}_{MB}=\frac{1}{4f^2}\langle\bar{B}i\gamma^\mu[(\Phi \partial_\mu\Phi - \partial_\mu\Phi\Phi))B - B(\Phi \partial_\mu\Phi - \partial_\mu\Phi\Phi))\rangle\label{LMB}
\end{equation}
where $f$ is the pion decay constant and the symbol $\langle\,\,\rangle$ denotes the trace in the flavor space of the $SU(3)$ matrices $\Phi$ and $B$ 
\begin{equation}
\Phi=\left(\begin{array}{ccc}\dfrac{1}{\sqrt{2}}\pi^0+\dfrac{1}{\sqrt{6}}\eta&\pi^+& K^+\\
\pi^-&-\dfrac{1}{\sqrt{2}}\pi^0+\dfrac{1}{\sqrt{6}}\eta&K^0\\
K^-&\bar{K}^0&-\dfrac{2}{\sqrt{6}}\eta\end{array}\right)\label{eqb}
\end{equation}
\\
\begin{equation}
B=\left(\begin{array}{ccc}\dfrac{1}{\sqrt{2}}\Sigma^0+\dfrac{1}{\sqrt{6}}\Lambda&\Sigma^+&p\\
\Sigma^-&-\dfrac{1}{\sqrt{2}}\Sigma^0+\dfrac{1}{\sqrt{6}}\Lambda&n\\
\Xi^-&\Xi^0&-\dfrac{2}{\sqrt{6}}\Lambda\end{array}\right)\label{phiBM}
\end{equation}

Following \cite{Inoue}, $\pi N$, $\eta N$, $K \Lambda$ and $K \Sigma$ are taken as the coupled channels for the pion-nucleon system.  For example, for total charge zero Eq. (\ref{bs}) is solved with the potential,
\begin{eqnarray}
v =
\left( \begin{array}{cccccc}
v_{\pi^0 n \rightarrow \pi^0 n} & v_{\pi^0 n \rightarrow \pi^- p} &  v_{\pi^0 n \rightarrow \eta n} & v_{\pi^0 n \rightarrow K^+ \Sigma^- } & v_{\pi^0 n \rightarrow K^0 \Sigma^0} & \cdots \\
v_{\pi^- p \rightarrow \pi^0 n} & v_{\pi^- p \rightarrow \pi^- p} &  v_{\pi^- p \rightarrow \eta n} & v_{\pi^- p \rightarrow K^+ \Sigma^-}   & v_{\pi^- p \rightarrow K^0 \Sigma^0 } & \cdots\\
v_{\eta  n\rightarrow \pi^0 n} & v_{\eta n \rightarrow \pi^- p} &  v_{ \eta n \rightarrow \eta n} & 
 \vdots & \vdots&\vdots\\
v_{K^+ \Sigma^- \rightarrow \pi^0 n} & v_{K^+ \Sigma^- \rightarrow \pi^- p} &  v_{K^+ \Sigma^- \rightarrow  \eta n} & \vdots & \vdots&\vdots\\
v_{ K^0 \Sigma^0 \rightarrow \pi^0 n} & v_{ K^0 \Sigma^0 \rightarrow \pi^- p} &  v_{ K^0 \Sigma^0 \rightarrow \eta n} & \vdots & \vdots&\vdots\\
v_{ K^0 \Lambda \rightarrow \pi^0 n} & v_{ K^0 \Lambda \rightarrow \pi^- p} &  v_{ K^0 \Lambda \rightarrow \eta n} & \vdots & \vdots&\vdots\\
\end{array} \right)\nonumber
\end{eqnarray}
and the $t_{\pi^0 n \rightarrow \pi^0 n}$ element of the resulting matrix is used in Eq. (\ref{tgt}) as $t^2$ and $t^1$. The  two body propagator, $\tilde{g}$ in Eq. (\ref{bs}), is divergent and is calculated using dimensional regularization
by taking the substraction constants from \cite{Inoue}, where the authors find the $N^*(1535)$ as a dynamically generated resonance in the $\pi N$ system and its coupled channels. 

\begin{figure}
\begin{center}
\includegraphics[scale=0.7]{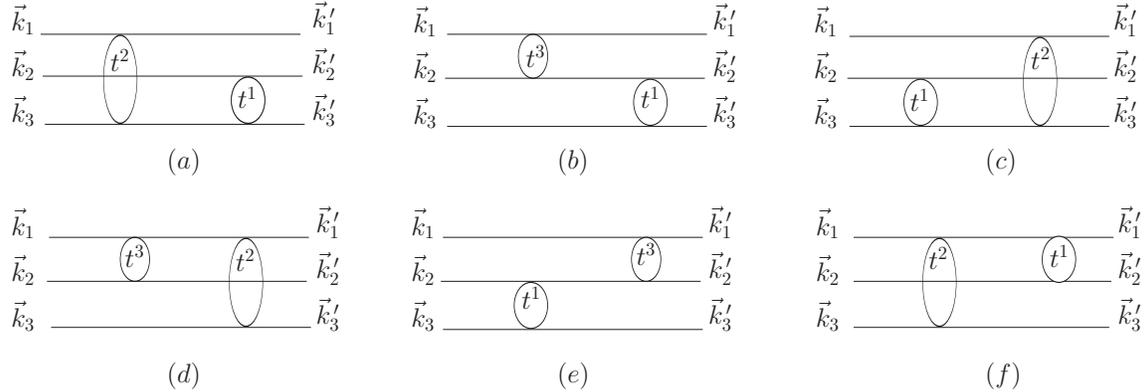}
\caption{\it Different possible diagrams including two successive interactions.}\label{alltgt}
\end{center}
\end{figure}
There are six possible three-body diagrams involving two $t$-matrices as shown in Fig. \ref{alltgt}. 
To calculate all these diagrams, we require the $\pi \pi$ $t$-matrices also, which have been obtained by solving the Bethe-Salpeter equation (Eq. (\ref{bs})) with $\pi \pi$, $\pi \eta$ and $K\bar{K}$ as coupled channels \cite{npa, oller}. The potentials for these channels have been calculated using the chiral Lagrangian \cite{weinberg, meissner, ecker, npa}
\begin{equation}
\mathcal{L}_{MM}=\dfrac{1}{12f^2}\langle(\partial_\mu\Phi\Phi-\Phi\partial_\mu\Phi)^2+M\Phi^4\rangle, \label{LMM}
\end{equation}
where
\begin{equation}
M=\left(\begin{array}{ccc}m_\pi^2&0&0\\0&m_\pi^2&0\\0&0&2m_K^2-m_\pi^2\end{array}\right)\nonumber
\end{equation}
and $m_\pi$, $m_K$ are the pion and kaon masses, respectively.
The two body propagator, $\tilde{g}$, in this case has also been calculated using the dimensional regularization \cite{oller}. A detailed study of these systems has been carried out in \cite{npa, oller} which revealed the dynamical generation of the $\sigma$ and $f_0$ resonances in the isospin zero sector and that of the $a_0$ in the isospin $1$ sector of these mesons. 

All the diagrams in Fig. \ref{alltgt} can be expressed mathematically as $t^i g^{ij} t^j$  with $i \ne j = 1, 2, 3$. In the above discussion we have taken one channel, $\pi^0 \pi^0 n$, as an example but the calculations have been carried out by taking five coupled channels into account. Hence  $t^i$, $g^{ij}$ and $t^j$ are matrices and each element of the $g^{ij}$ matrix is given by
\begin{equation}
g^{ij} (\vec{k_i}^\prime, \vec{k_j}) = \Biggr( \prod_{r=1}^D 
 \frac{N_r}{2 E_r} \Biggr) \frac{1}{\sqrt{s} - E_i (\vec{k_i}^\prime) - E_l(\vec{k_i}^\prime + \vec{k_j}) - E_j(\vec{k_j})}, \vspace{1cm}  l \ne i, l \ne j,=1,2,3 \label{gij}
\end{equation}
where $D$ is the number of particles propagating between two $t$-matrices. Following the normalization of \cite{mandlshaw}, $N_r = 1$ for a meson and $N_r = 2 M_r$ for a baryon with $M_r$ being the mass of the baryon and $\vec{k_i}^\prime (\vec{k_j})$ is the momentum of the $i$th ($j$th) particle in the final (initial) state.

\subsection{Kinematics}
We now define the kinematics for the system. There are two variables in the calculation; the total energy of the three-body system, denoted as $\sqrt{s}$, and the invariant mass of the particles $2$ and $3$, denoted as $\sqrt{s_{23}}$. The other invariant masses are obtained in terms of these variables as
\begin{equation}
s_{ij} = s + m_k^2 -\frac{\sqrt{s} (\sqrt{s} - E_1) (s_{23} + m_k^2 - m_j^2)}{s_{23}}\label{sij}
\end{equation}
with $m_k$ being the mass of the non-interacting particle and
\begin{equation}
E_1 = \dfrac{s - s_{23} + m_1^2 }{2 \sqrt{s}}.
\end{equation}
The definition in Eq. (\ref{sij}) implies an angular average between external momenta suited for the study of s-waves.

From all this we can calculate the momenta, $|\vec{k}_1|, |\vec{k}_1^\prime|$ of the particle $1$ in the global center of mass system and that of the particles $2$ and $3$ in their rest frame ($R_{23}$), which we denote as $\vec{K} \,(\vec{K}^\prime)$ in the initial (final) state;
\begin{eqnarray}
&&|\vec{k}_1| = |\vec{k}_1^\prime| = \dfrac{1}{2\sqrt{s}}\lambda^{1/2}(s,s_{23},m_1^2)\\ \nonumber
&&|\vec{K}| = |\vec{K}^\prime| = \dfrac{1}{2\sqrt{s_{23}}}\lambda^{1/2}(s_{23},m_2^2,m_3^2).
\end{eqnarray}
The calculation of the $g^{ij}$ propagators for different diagrams requires the momenta of the particles in the global center of mass. For this, we boost the momentum in $R_{23}$ to the global center of mass using the relations \cite{Fernandez}:
\begin{eqnarray}
\vec{k}_2&=&\Bigg[\Big(\dfrac{\sqrt{s}-E_1(\vec{k}_1)}{\sqrt{s_{23}}}-1\Big)\dfrac{\vec{K}\cdot\vec{k}_1}
{{\vec{k}_1}^2}-\dfrac{E_2^{R_{23}}(\vec{K})}{\sqrt{s_{23}}}\Bigg]\vec{k}_1+\vec{K}\nonumber\\
\vec{k}_3&=&\Bigg[\Big(\dfrac{\sqrt{s}-E_1(\vec{k}_1)}{\sqrt{s_{23}}}-1\Big)\dfrac{(-\vec{K})\cdot\vec{k}_1}{{\vec{k}_1}^2}-\dfrac{E_3^{R_{23}}(\vec{K})}{\sqrt{s_{23}}}\Bigg]\vec{k}_1-\vec{K}\nonumber\\
{\vec{k}_2}^\prime&=&\Bigg[\Big(\dfrac{\sqrt{s}-E_1({\vec{k}_1}^\prime)}{\sqrt{s_{23}}}-1\Big)
\dfrac{\vec{K}^\prime\cdot{\vec{k}_1}^\prime}{\vec{k}_1^{\prime \,2}}-\dfrac{{E_2}^{R_{23}}(\vec{K}^\prime)}{\sqrt{s_{23}}}\Bigg]{\vec{k}_1}^\prime+\vec{K}^\prime\nonumber\\
{\vec{k}_3}^\prime&=&\Bigg[\Big(\dfrac{\sqrt{s}-E_1({\vec{k}_1}^\prime)}{\sqrt{s_{23}}}-1\Big)\dfrac{(-\vec{K}^\prime)
\cdot{\vec{k}_1}^\prime} {{\vec{k}_1}^{\prime\,2}} - \dfrac{{{E_3}^{R_{23}}} (\vec{K}^\prime)} {\sqrt{s_{23}}}\Bigg] \vec{k}^{\,\prime}_1 - \vec{K}^\prime.
\end{eqnarray}
We define  $\vec{k}_1$ to be along the z-axis and  $\vec{K}$ to form a plane with $\vec{k}_1$, i.e.,
\begin{displaymath}
\begin{array}{cc}
\vec{k}_1 = \left\{\begin{array}{lll} 0\\ 0\\ |\vec{k}_1|\\ \end{array} \right\} & \vec{K} = \left\{ \begin{array}{lll} |\vec{K}| sin(\theta_K)\\ 0\\ |\vec{K}| cos(\theta_K)\\ \end{array} \right\}.
\end{array}
\end{displaymath}
 
The $t$-matrices are calculated as a function of the invariant mass of the interacting particles, e.g., $t^1$ is calculated as a function of $\sqrt{s_{23}}$. This is so because, in the chiral approach, the $t$-matrices can be split into an off-shell part, which behaves as $(q_i^2 - m_i^2)$, with $q_i$ being the four vector of the off-shell particle, and an on-shell part, where $q_i^2$ is set to $m_i^2$. In analogy to the findings in \cite{mko1}, this off-shell dependence of the $t$-matrices in the three-body diagrams is found to cancel exactly with the three-body forces generated from the chiral Lagrangian for 2 meson + baryon $\rightarrow$ 2 meson + baryon contact term in the SU(2) limit (see the Appendix).  In \cite{mko3}, where a study of the $\phi K \bar{K}$ system has been carried out and dynamical generation of the $X(2175)$ resonance is found, an explicit calculation including the off-shell parts of the $t$-matrices has been done. The results obtained in this case are found to be qualitatively similar to those obtained by implementing the cancellation of the off-shell parts of the $t$-matrices with the chiral three body forces. The only difference in the results was a shift in the peak position in the squared amplitude by 40 MeV (about 2 $\%$ of the mass of the resonance). However it should be noted that the $t$-matrices and hence their off-shell parts, which have been derived from a chiral Lagrangian, are representation dependent. In the present work we rely upon the cancellation discussed above and retain the on-shell parts of the $t$-matrices, which depend on the invariant mass of the interacting pair in s-wave (as considered here).

\subsection{Higher order diagrams}

\begin{figure}
\begin{center}
\includegraphics[scale=0.7]{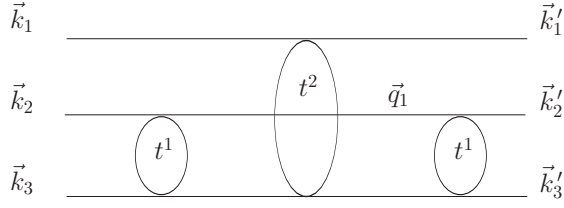}
\caption{\it A diagram involving three $t$-matrices.}\label{tgtgt1}
\end{center}
\end{figure}
The calculation of the diagrams with more than two $t$-matrices involve a loop of three propagators and three two-body $t$-matrices. Such diagrams can be written as $t^i g^{ij} t^j g^{jl} t^{l}$, where the propagators and the $t^j$-matrix depend on the loop variable even though the $t$-matrices in our model are calculated as a function of the invariant mass of the interacting pair. The calculation of the three-body scattering matrix would simplify if we could extract $t^j g^{jl} t^{l}$ out of the loop integral. However, we would like to keep the loop dependence of this term. This can be done if the $t^j g^{jl}$ calculated with off-shell variables is absorbed in the previous propagator as in \cite{mko1}. Following the formalism developed in \cite{mko1}, we write the diagram shown in Fig. \ref{tgtgt1} as 
\begin{equation}
t^1 G^{121} t^2 g^{21} t^1 = t^1 (\sqrt{s_{23}}) G^{121} t^2 (\sqrt{s_{13}}) g^{21} (\vec{k}_2^\prime, \vec{k}_1) t^1 (\sqrt{s_{23}}), \label{tGtgt}
\end{equation}
where
\begin{equation}
G^{121} = \int \frac{d\vec{q}_1}{(2 \pi)^3} \frac{1}{2 E_2 (\vec{q}_1)} \frac{ M_3}{ E_3 (\vec{q}_1)} \frac{1}{\sqrt{s_{23}} - E_2 (\vec{q}_1) - E_3 (\vec{q}_1) + i\epsilon} \times F^{121} (\vec{q}_1, \vec{k}_2^\prime, \vec{k}_1, s_{13} ) \label{g313}
\end{equation}
with
\begin{equation}
F^{121} (\vec{q}_1, \vec{k}_2^\prime, \vec{k}_1, s_{13} ) = t^2( s_{13}^{q_1} ) \times g^{21} (\vec{q}_{1}, \vec{k}_1) \times [g^{21} (\vec{k}_2^\prime, \vec{k}_1)]^{-1} \times [ t^2  (\sqrt{s_{13}} )]^{-1}.
\end{equation}

Note that, while $s_{23}$ is defined from the external variables for the diagram shown in Fig. \ref{tgtgt1}, the argument $s_{13}$ of the $t^2$-matrix is a function of the loop variable and must be kept in the loop integral. We, thus, introduce $t^2$ calculated as a function of 
\begin{equation}
s_{13}^{q_1} = s - m_2^2 - 2 \sqrt{s} \,\frac{E_2( \vec{q}_{1}) (\sqrt{s} - E_3(\vec{k}_3^\prime))}{\sqrt{s_{12}}} \label{sijoff}
\end{equation}
in  $F^{121}$ and hence in the loop integral $G^{121}$. $F^{121}$ also contains the inverse of $t^2$ calculated as a
function of $s_{13}$ evaluated in terms of on-shell variables (Eq. (\ref{sij})) and the $g^{21}$ propagator depending on off-shell variables along with the inverse of its on-shell version. 

In this way,  $[g^{21} (\vec{k}_2^\prime, \vec{k}_1)]^{-1} \times [ t^2  (\sqrt{s_{13}} )]^{-1}$ in $F^{121}$ (and hence in $G^{121}$) when multiplied to $t^2(\sqrt{s_{13}}) g^{21} (\vec{k}_2^\prime, \vec{k}_1)$ in Eq. (\ref{tGtgt}) give an identity leaving $t^2 g^{21}$ evaluated with the loop variable in the loop integral. Simplifying Eq. (\ref{tGtgt}) we have
\begin{equation}
t^1 (\sqrt{s_{23}})   \int \frac{d\vec{q}_1} {(2 \pi)^3} \frac{1}{2 E_2 (\vec{q}_1)} \frac{ M_3}{ E_3 (\vec{q}_1)} \frac{1}{\sqrt{s_{23}} - E_2 (\vec{q}_1) - E_3 (\vec{q}_1) + i\epsilon}  t^2(s_{13}^{q_1})  g^{21} (\vec{q}_1, \vec{k}_1)  t^1 (\sqrt{s_{23}})
\end{equation}
which is the right contribution of the diagram in Fig. \ref{tgtgt1}. The integrals of $G^{ijk}$  are regularized with a cut-off of 1 GeV in the modulus of the momentum, which if changed to 1.5 Gev introduces less than $1\%$ of a change in $G$. One of the propagators in the $G^{121}$ function is evaluated in the center of mass frame of two particles for convenience. A diagram with three $t$-matrices is, thus, written as $t^iG^{ijk}t^jg^{jk}t^k$ instead of $t^ig^{ij}t^jg^{jk}t^k$.
As has been discussed in \cite{mko1}, for diagrams with three $t$-matrices in general and in \cite{mko2} for a specific diagram in detail, this scheme simplifies the calculations. The formalism is further developed by repeating the above procedure for higher order diagrams too, i.e., by replacing the $g^{ij}$ propagator by the $G^{ijk}$ loop function everytime a new interaction is added.

\begin{figure}
\begin{center}
\includegraphics[scale=0.7]{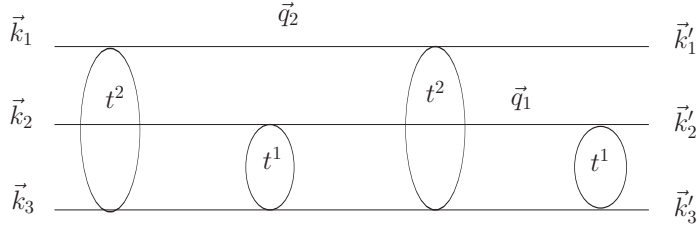}
\caption{\it  A diagram with two concatenated loops.} \label{4ts}
\end{center}
\end{figure}

In case of the diagrams with more than three $t$-matrices, this scheme involves an approximation, since the introduction of a new interaction, to the diagrams of the kind shown in Fig. \ref{tgtgt1}, replaces the external variables in the latter case by variables of a former loop. This procedure, which renders the integral Faddeev equations into a set of algebraic equations, is certainly very economical in terms of numerical solution and a justification for its use is given below.  

\begin{figure}
\begin{center}
\includegraphics[scale=0.4]{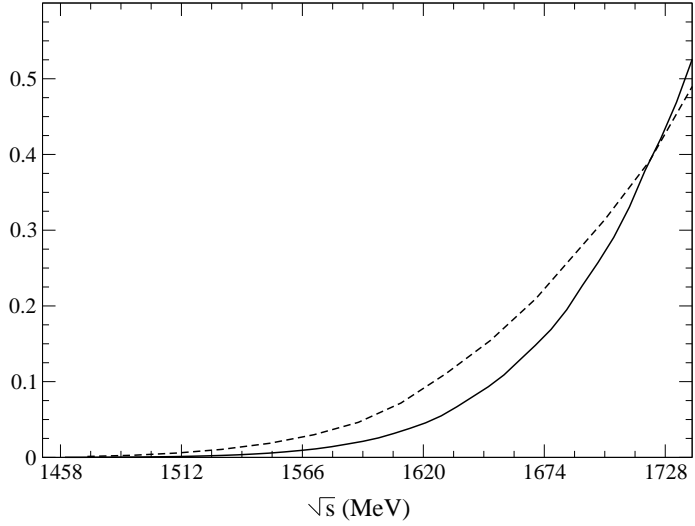}
\caption{\it The comparison of modulus square of Eq. (\ref{our}) and Eq. (\ref{orig}) shown as dashed and solid lines, 
respectively, in units of $10^{-15}$ MeV$^{-6}$.}\label{t14}
\end{center}
\end{figure}

Let us discuss in detail a diagram with four $t$-matrices as shown in Fig. \ref{4ts}, as an example.
This diagram is written explicitly as 
\begin{eqnarray}
&&t^1 G^{121} t^2 G^{212} t^1 g^{12} t^2 =
t^1( \sqrt{s_{23}} ) 
\Biggr( \int \frac{d\vec{q}_1}{(2\pi)^3} \frac{1}{2 E_2 (\vec{q}_1)} \frac{ M_3}{ E_3 (\vec{q}_1)} \frac{1}{\sqrt{s_{23}} - E_2 (\vec{q}_1) - E_3 (\vec{q}_1) + i\epsilon} \nonumber \\ 
&&\times t^2 ( s_{13}^{q_1} ) \, g^{21}(\vec{q}_1, \vec{k}_1) \,[g^{21}(\vec{k}_2^\prime, \vec{k}_1)]^{-1} \, [t^2( \sqrt{s_{13}} )]^{-1} \Biggr) 
t^2( \sqrt{s_{13}} ) \nonumber \\
&&\Biggr( \int \frac{d\vec{q}_2}{(2\pi)^3} \frac{1}{2 E_1 (\vec{q}_2)} \frac{ M_3}{ E_3 (\vec{q}_2)} \frac{1}{\sqrt{s_{13}} - E_1 (\vec{q}_2) - E_3 (\vec{q}_2) + i\epsilon} t^1( s_{23}^{q_2} ) \,\,g^{12}(\vec{q}_2, \vec{k}_2) \, [g^{12}(\vec{k}_1^\prime, \vec{k}_2)]^{-1} \nonumber \\ 
&& \times [t^1( \sqrt{s_{23}} )]^{-1}\Biggr)\, t^1( \sqrt{s_{23}} ) \, g^{12}(\vec{k}_1^\prime, \vec{k}_2) \, t^2( \sqrt{s_{13}} )\label{our2}
\end{eqnarray}
which can be simplified to
\begin{eqnarray}
&&
t^1( \sqrt{s_{23}} ) 
\int \frac{d\vec{q}_1}{(2\pi)^3} \frac{1}{2 E_2 (\vec{q}_1)} \frac{ M_3}{ E_3 (\vec{q}_1)} \frac{1}{\sqrt{s_{23}} - E_2 (\vec{q}_1) - E_3 (\vec{q}_1)  + i\epsilon} t^2 ( s_{13}^{q_1} )  g^{21}(\vec{q}_1, \vec{k}_1)   \nonumber \\ 
&&\times
[g^{21}(\vec{k}_2^\prime, \vec{k}_1)]^{-1}  \int \frac{d\vec{q}_2}{(2\pi)^3} \frac{1}{2 E_1 (\vec{q}_2)} \frac{ M_3}{ E_3 (\vec{q}_2)} \frac{1}{\sqrt{s_{13}} - E_1 (\vec{q}_2) - E_3 (\vec{q}_2)  + i\epsilon}
t^1(  s_{23}^{q_2} ) \nonumber \\ 
&&\times g^{12}(\vec{q}_2, \vec{k}_2)  t^2( \sqrt{s_{13}} ),
\label{our}
\end{eqnarray}
where $s_{23}^{q_2}$ is calculated analogously to Eq.(\ref{sijoff}).
We compare our expression (\ref{our}) with the corresponding one written in terms of the $g$ propagators of the 
concatenated two loops
\begin{equation}
t^1( \sqrt{s_{23}} ) 
\Biggr[ \int \frac{d\vec{q}_1}{(2\pi)^3} \int \frac{d\vec{q}_2}{(2\pi)^3} g^{12}(\vec{k}_1^\prime, \vec{q}_1) t^2 ( s_{13}^{q_1}) g^{21}(\vec{q}_1, \vec{q}_2) t^1 ( s_{23}^{q_2} ) g^{12}(\vec{q}_2,\vec{k}_2) \Biggr] t^2( \sqrt{s_{13}}).
\label{orig}
\end{equation}
The dependence of $g^{21}$ on the two loop variables has been thus factorized in Eq. (\ref{our}) as
\begin{equation}
g^{21}(\vec{q}_1, \vec{q}_2) = \mathscr{F}_1 (\vec{q}_1)  \mathscr{F}_2 (\vec{q}_2),\label{factorize}
\end{equation}
where
\begin{equation}
\mathscr{F}_1 (\vec{q}_1) = g^{21}(\vec{q}_1, \vec{k}_1) [g^{21}(\vec{k}_2^\prime, \vec{k}_1)]^{-1}
\end{equation}
and
\begin{equation}
 \mathscr{F}_2 (\vec{q}_2) = g^{21}(\vec{q}_2) = \frac{1}{2 E_1 (\vec{q}_2)} \frac{ M_3}{ E_3 (\vec{q}_2)} \frac{1}{\sqrt{s_{13}} - E_1 (\vec{q}_2) - E_3 (\vec{q}_2)}.
\end{equation}
This factorization, which simplifies the calculations to a great extent, leads to very similar results to those obtained with the concatenated loop function as can be seen in Fig. \ref{t14}, where we show the mod-square of the Eq. (\ref{our}) and Eq. (\ref{orig}) as a function of $\sqrt{s}$, in the energy region of our interest.
The agreement of the results depicted in Fig. \ref{t14} shows that Eq. (\ref{our}) is a good approximation of the Eq. (\ref{orig}). Hence, this scheme is used to write the rest of the higher order diagrams which contribute to the three-body amplitude. 

If we sum Eqs. (\ref{tgt1}), (\ref{tGtgt}), (\ref{our2}) and all the other possible diagrams with the last two
$t$-matrices as $t^2$ and $t^1$, we get the series
\begin{equation}
t^1g^{12}t^2 + t^1G^{121}t^2g^{21}t^1 + t^1G^{121}t^2G^{212}t^1g^{12}t^2 +  \cdots + t^1G^{123}t^2g^{23}t^3 + t^1G^{123}t^2G^{232}t^3g^{32}t^2 + \cdots,
\end{equation}
which we define as $T_R^{12}$. Similarly, we consider all other possible diagrams obtained by permutating different interactions between the three hadrons and get the following equations upon summing all the diagrams with the same last two $t$-matrices
\begin{eqnarray} \nonumber
T^{\,12}_R&=&t^1g^{12}t^2+t^1\Big[G^{\,121\,}T^{\,21}_R+G^{\,123\,}T^{\,23}_R\Big] \\ \nonumber
T^{\,13}_R&=&t^1g^{13}t^3+t^1\Big[G^{\,131\,}T^{\,31}_R+G^{\,132\,}T^{\,32}_R\Big] \\ \nonumber
T^{\,21}_R&=&t^2g^{21}t^1+t^2\Big[G^{\,212\,}T^{\,12}_R+G^{\,213\,}T^{\,13}_R\Big] \\ \nonumber
T^{\,23}_R&=&t^2g^{23}t^3+t^2\Big[G^{\,231\,}T^{\,31}_R+G^{\,232\,}T^{\,32}_R\Big] \\ \nonumber
T^{\,31}_R&=&t^3g^{31}t^1+t^3\Big[G^{\,312\,}T^{\,12}_R+G^{\,313\,}T^{\,13}_R\Big] \\ 
T^{\,32}_R&=&t^3g^{32}t^2+t^3\Big[G^{\,321\,}T^{\,21}_R+G^{\,323\,}T^{\,23}_R\Big] 
\end{eqnarray}\label{trest}
These are six coupled equations which are summed to get
\begin{equation}
T_R = T_R^{12} + T_R^{13} + T_R^{21} + T_R^{23} + T_R^{31} + T_R^{32}.\label{ourfullt}
\end{equation}
The $T_R^{ij}$ can be related to the Faddeev partitions $T^i$ of the 
the Faddeev equations
\begin{equation}
T=T^1+T^2+T^3\label{fullt}
\end{equation}
as 
\begin{equation}
T^i =t^i\delta^3(\vec{k}^{\,\prime}_i-\vec{k}_i) + T_R^{ij} + T_R^{ik}\label{Ti}.
\end{equation}

\section{Results and discussions}
The $T_R$ in Eq. (\ref{ourfullt}) has been calculated in s-wave for the coupled channels $\pi^0 \pi^0 n$, $\pi^+ \pi^- n$, $\pi^- \pi^+ n$, $\pi^0 \pi^- p$, $\pi^- \pi^0 p$ as a function of $\sqrt{s}$ and $\sqrt{s_{23}}$. All the angle dependent expressions have thus been projected into $s$-wave. The $T_R$-matrix (Eq. (\ref{ourfullt})) is then projected on the isospin base defined in terms of the total isospin of the three body system, $I$, and the total isospin of two pions, $I_{\pi\pi}$, defining the states 
as $|I, I_{\pi\pi} \rangle$. These states are obtained assuming the phase convention for  $\mid \pi^+\rangle$ as $-\mid 1,1\rangle$. We write the state $\mid \pi^0 \,\pi^0 \,n \rangle$, for example, as
\begin{eqnarray}
\mid \pi^0\,\pi^0\,n\rangle&=&\mid 1, 0 \rangle \otimes \mid 1, 0 \rangle \otimes \mid 1/2, -1/2 \rangle\nonumber\\
&=&\left\{\sqrt{\frac{2}{3}}\mid I_{\pi\pi} = 2, I_{\pi\pi}^z = 0 \rangle - \sqrt{\frac{1}{3}}\mid I_{\pi\pi} = 0, I_{\pi\pi}^z = 0 \rangle \right\} \otimes
\nonumber\\&&\otimes \mid 1/2, -1/2 \rangle\nonumber\\
&=&\sqrt{\frac{2}{5}}\mid I = 5/2, I_{\pi\pi} = 2\rangle+\frac{2}{\sqrt{15}}\mid I = 3/2,  I_{\pi\pi} = 2 \rangle - \sqrt{\frac{1}{3}}\mid I = 1/2,  I_{\pi\pi} = 0 \rangle\nonumber
\end{eqnarray}
To simplify the notation, we omit the label $I$ and $I_{\pi\pi}$ and write
\begin{equation}
\mid \pi^0\,\pi^0\,n\rangle = \sqrt{\frac{2}{5}}\mid 5/2, 2 \rangle+\frac{2}{\sqrt{15}}\mid 3/2, 2\rangle - \sqrt{\frac{1}{3}}\mid 1/2, 0 \rangle.
\end{equation}
Similarly,
\begin{eqnarray}
\mid \pi^+\, \pi^- \,n \rangle&=&-\sqrt{\frac{1}{10}}\mid 5/2, 2 \rangle - \sqrt{\frac{1}{15}}\mid 3/2, 2\rangle-
\sqrt{\frac{1}{3}}\mid3/2, 1 \rangle - \sqrt{\dfrac{1}{6}} \mid 1/2, 1 \rangle - \sqrt{\frac{1}{3}}\mid 1/2, 0 \rangle \nonumber\\
\mid \pi^-\,\pi^+\, n \rangle&=&-\sqrt{\frac{1}{10}}\mid 5/2, 2\rangle-\sqrt{\frac{1}{15}}\mid 3/2, 2\rangle+
\sqrt{\frac{1}{3}}\mid3/2, 1 \rangle + \sqrt{\dfrac{1}{6}} \mid 1/2, 1 \rangle - \sqrt{\frac{1}{3}}\mid 1/2, 0 \rangle
\nonumber\\
\mid \pi^-\,\pi^0\,p \rangle&=&\sqrt{\frac{1}{5}}\mid 5/2, 2 \rangle - \sqrt{\frac{3}{10}}\mid 3/2, 2 \rangle -
\sqrt{\frac{1}{6}}\mid 3/2, 1 \rangle +\sqrt{\frac{1}{3}} \mid 1/2, 1 \rangle
\nonumber\\
\mid \pi^0\,\pi^-\,p \rangle&=&\sqrt{\frac{1}{5}}\mid 5/2, 2 \rangle - \sqrt{\frac{3}{10}}\mid 3/2, 2 \rangle +
\sqrt{\frac{1}{6}}\mid 3/2, 1 \rangle -\sqrt{\frac{1}{3}} \mid 1/2, 1 \rangle. \label{isos}
\end{eqnarray}
From Eqs. (\ref{isos}), one can obtain 
\begin{eqnarray}\nonumber
\mid 5/2, 2 \rangle&=& \sqrt{\frac{1}{5}} \biggr( \sqrt{2} \mid \pi^0\,\pi^0\,n\rangle + \mid \pi^0\,\pi^-\,p \rangle + \mid \pi^-\,\pi^0\,p \rangle -\sqrt{\frac{1}{2}} \mid \pi^+\, \pi^- \,n \rangle - \sqrt{\frac{1}{2}} \mid \pi^-\, \pi^+ \,n \rangle \biggr)\\ \nonumber
\mid 3/2, 2 \rangle&=& \sqrt{\frac{1}{15}} \biggr( 2 \mid \pi^0\,\pi^0\,n\rangle - \frac{3}{\sqrt{2}} \mid \pi^0\,\pi^-\,p \rangle -  \frac{3}{\sqrt{2}} \mid \pi^-\,\pi^0\,p \rangle - \mid \pi^+\, \pi^- \,n \rangle - \mid \pi^-\, \pi^+ \,n \rangle \biggr)\\ \nonumber
\mid 1/2, 0 \rangle&=& -\sqrt{\frac{1}{3}} \biggr( \mid \pi^0\,\pi^0\,n\rangle + \mid \pi^+\, \pi^- \,n \rangle +  \mid \pi^-\, \pi^+ \,n \rangle \biggr).
\end{eqnarray}

One could equivalently define the states in terms of the total isospin and the isospin of a pion-nucleon sub-system ($I_{\pi N})$ by repeating the former procedure or using the Racah coefficients for the transformation of the $\mid I, I_{\pi\pi} \rangle$ states to $\mid I, I_{\pi N} \rangle$ states. 

\begin{figure}
\begin{center}
\includegraphics[width=0.8\textwidth]{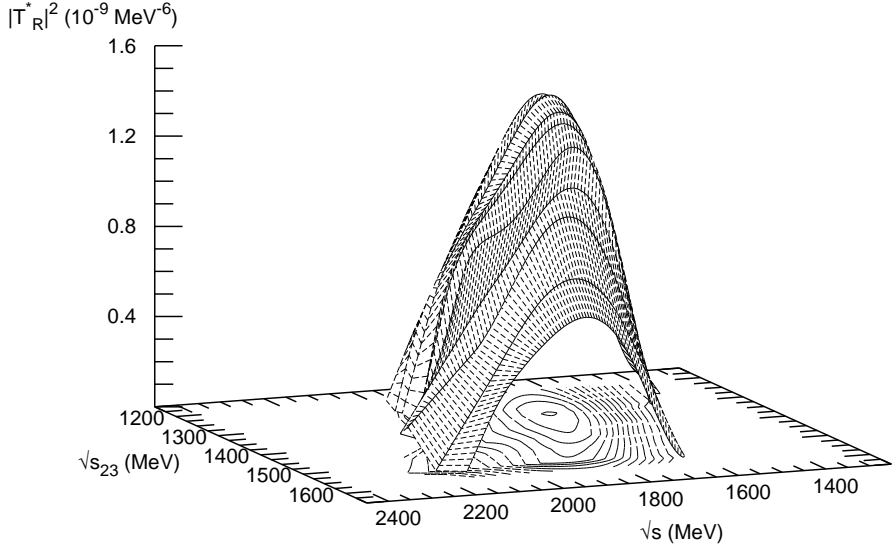}
\caption{\it The squared amplitude for the $\pi \pi N$ system in isospin 1/2 configuration as a function of $\sqrt{s}$ and $\sqrt{s_{23}}$.}\label{1710_1}
\end{center}
\end{figure}

\begin{figure}
\begin{center}
\includegraphics[angle=-90, width=0.6\textwidth]{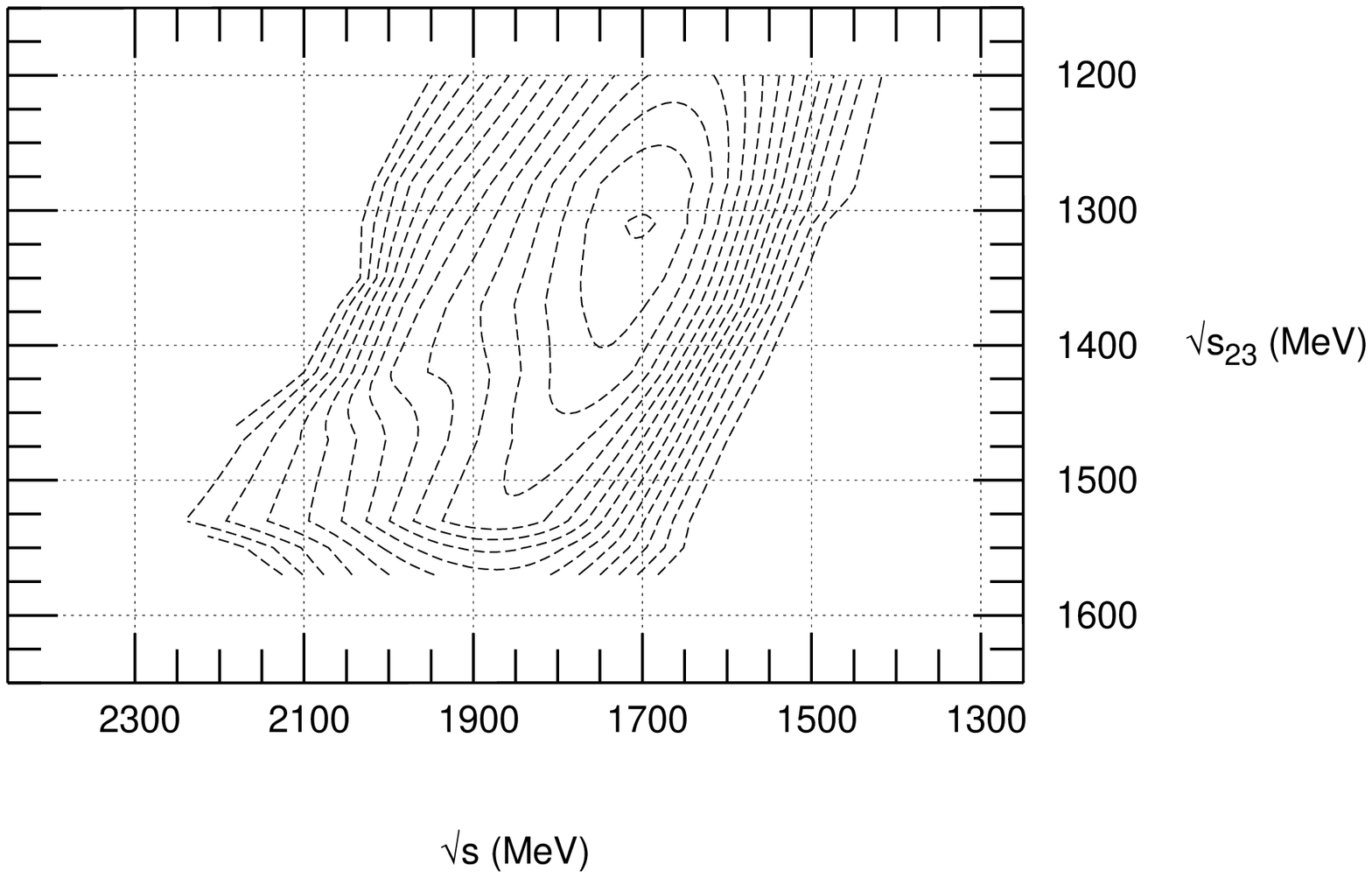}
\caption{\it The projection of the amplitude shown in Fig. \ref{1710_1}.}\label{1710_3}
\end{center}
\end{figure}

\begin{figure}
\begin{center}
\includegraphics[angle=-90, width=0.6\textwidth]{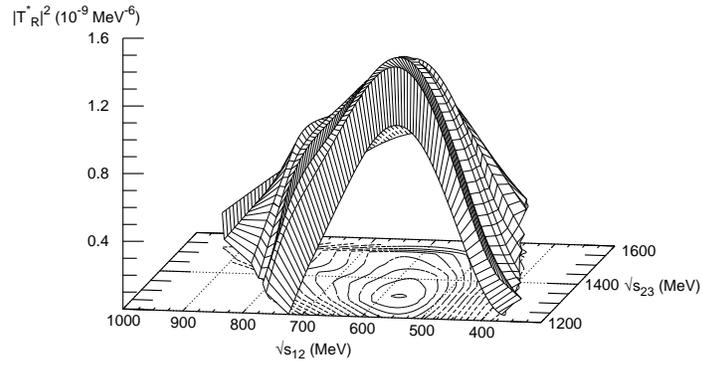}
\caption{\it The same as shown in Fig. \ref{1710_1} but as a function of the $\pi \pi$ invariant mass and that of the $\pi N$ system.}\label{1710_2}
\end{center}
\end{figure}

\begin{figure}
\begin{center}
\includegraphics[angle=-90, width=0.6\textwidth]{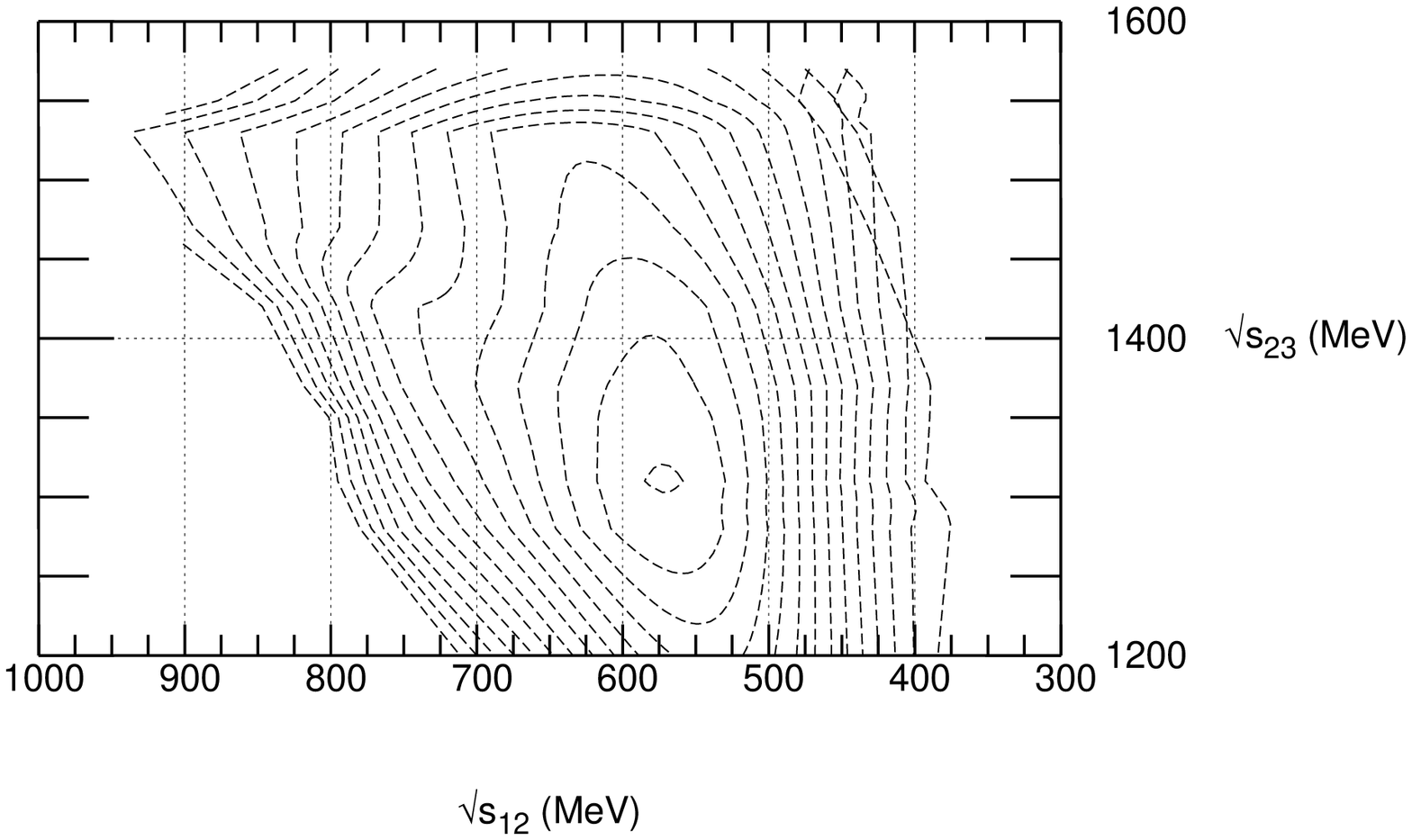}
\caption{\it The projection of the amplitude shown in Fig. \ref{1710_2}}\label{1710_4}
\end{center}
\end{figure}

In Fig. \ref{1710_1} we show the squared amplitude $\mid T_R^* \mid^2 = \mid T_R -\sum\limits_{i \ne j = 1}^3 t^i g^{ij} t^j\mid^2$ for the $\pi\pi N$ system, calculated in s-wave and projected on the isospin base $\mid I, I_{\pi\pi} \rangle = \mid 1/2, 0 \rangle$. The $\sum tgt$ has been subtracted out of the $T_R$ (following \cite{mko1}) since it does not give rise to any three-body structure and only provides a background to the amplitude. The squared amplitude shown in Fig. \ref{1710_1} has a peak at $\sqrt{s} = 1704$ MeV, with a full width at half maximum of 375 MeV (see also Fig. \ref{1710_3}). These results are in good agreement with the characteristics of the $N^*(1710)$ \cite{pdg} and, hence, we relate the resonance shown in Fig. \ref{1710_1} with the $N^*(1710)$. To get further physical meaning of this peak, we show the same amplitude depicted in Fig. \ref{1710_1}, but as a function of $\sqrt{s_{23}}$ and $\sqrt{s_{12}}$ in Fig. \ref{1710_2}. The peak in $\sqrt{s_{12}}$ is very wide (width $\sim$ 270 MeV ) and is in the energy region of the $\sigma$ resonance (see also Fig. \ref{1710_4}). This means that the $N^*(1710)$ has a large $\pi\pi N$ component where the $\pi\pi$ sub-system rearranges itself as the $\sigma$ resonance.

Although we find evidence for the $N^*(1710)$, this work fails to find any clear trace of the  Roper resonance, which means that
considering the $\pi\pi N$ system in s-wave interaction does not suffice to generate the Roper resonance, which is not surprising. Other works such as the Juelich model \cite{juelich}, which successfully describes the dynamical generation of the Roper resonance, contains additional information on the $\pi N$, $\pi \Delta, \rho N$ coupled channels and $\sigma N$ forces beyond the three body contact term of the chiral Lagrangians which we include here and which cancels the off-shell dependence of the amplitudes. An important contribution of the $\pi \Delta$ channel and  $\pi \pi$ final state interaction (with one of the pions coming from the decay of the $\Delta$ resonance) to the Roper resonance has also been claimed in \cite{eli}. Such information is not present in our formalism. Things are different in the case of the  $N^*(1710)$ with its large empirical coupling to  $\pi \pi N$ and weaker to  $\pi N$  and other coupled channels.

Another important result of this work is that we do not find any resonant structure in the total isospin $I = 3/2$ and $I = 5/2$ configuration. Should we have found the latter, it would be exotic in the sense that it would not be possible to construct it with just three quarks. But no structure is found in this isospin state.

To summarize, we have studied the $\pi\pi N$ system in s-wave, thus in $J^\pi = 1/2^+$ configuration. We find a resonance, in three-dimensional plots of the squared amplitude versus the total energy and the invariant mass of a sub-system, at 1704 MeV, which can be associated with the $N^*(1710)$ \cite{pdg}. Our peak has a full width $\Gamma$ = 375 MeV to be compared with that of the $N^*(1710)$ which ranges from 90-500 MeV  \cite{pdg}. We find that the invariant mass of the $\pi\pi$ sub-system falls in the region of the mass of the $\sigma$ (500 -i 200 MeV) when the $\pi\pi N$ amplitude peaks at $\sqrt{s} = $ 1704 MeV, which means that the large width of the $N^*(1710)$ could be related to that of the $\sigma$ resonance formed in the $\pi\pi$ sub-system. No evidence for states with $I = 3/2$ and $I = 5/2$  is found in this work. We also do not find the Roper resonance in our approach. This should not be seen as a negative result, but as an evidence that the structure of the Roper is far more complex than that envisaged by the $\pi \pi N$ interaction in s-wave, which is what we have investigated in the present work.

\section{Acknowledgments}
This work is partly supported by
DGICYT contract number FIS2006-03438,
and the Generalitat Valenciana. A. M. T. wishes to acknowledge support from a FPU
fellowship of the
Ministerio de Educaci\'on y Ciencia.
This research is  part of
the EU Integrated Infrastructure Initiative  Hadron Physics Project under
contract number RII3-CT-2004-506078, and K. P. K. wishes to acknowledge direct support from it. 

\vspace{1cm}
\begin{center}
{\bf \Large Appendix}
\end{center}

In this appendix we discuss the three-body interactions including the off-shell parts of the $t$-matrices that give rise to a kind of three body force which, as we  show below, gets cancelled with the three-body force arising from the chiral Lagrangian. 

\begin{figure}
\begin{center}
\includegraphics[width=0.7\textwidth]{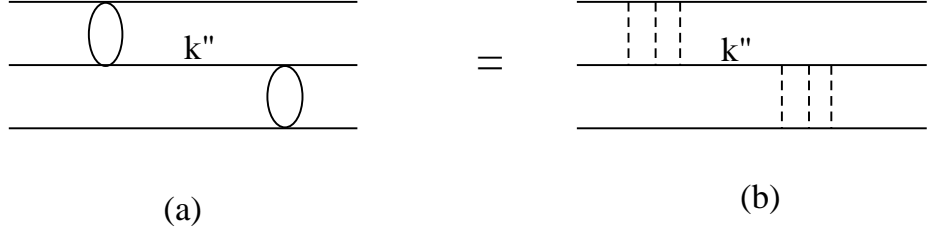}
\caption{\it A diagrammatic representation of the $t^1 g^{13} t^3$ term. The blob in (a) represents a $t$-matrix which can be expressed mathematically as $v + v\tilde{g}v + v\tilde{g}v\tilde{g}v  +  v\tilde{g}v\tilde{g}v\tilde{g}v + ...$. And (b) shows the term  $(v^1 \tilde{g}^1 v^1 \tilde{g}^1 v^1) g^{13} (v^3 \tilde{g}^3 v^3 \tilde{g}^3 v^3 ) $ of Eq. (\ref{vgv}).}\label{3b}
\end{center}
\end{figure}

Let us consider the lowest order diagrams, which correspond to the first terms of the $T_R$ equations, i.e., $t^i g^{ij}t^j$.  There are six terms of this kind shown in the Fig. \ref{alltgt}, which can be expanded in terms of the potentials as
\begin{eqnarray}
t^i g^{ij}t^j &=& \big[ v^i + v^i \tilde{g}^i v^i + v^i \tilde{g}^i v^i \tilde{g}^i v^i + \cdots \big] \, g^{ij} \, \big[v^j + v^j \tilde{g}^j v^j + v^j \tilde{g}^j v^j \tilde{g}^j v^j + \cdots \big] \nonumber \\
&=& v^i g^{ij} v^j + v^i \tilde{g}^i  v^i g^{ij} v^j  + v^i  g^{ij} v^j \tilde{g}^j v^j + \cdots \label{vgv}
\end{eqnarray}
For example, a term of $t^1 g^{13} t^3$ expanded as in Eq. (\ref{vgv}) is shown in Fig. \ref{3b}.

\begin{figure}
\centering
\includegraphics[width=0.3\textwidth]{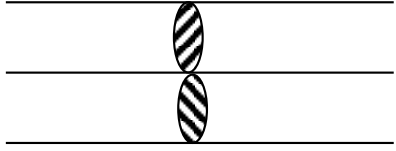}
\caption{\it An induced effective three-body force generated by the cancellation of the off-shell part of the potential and a propagator as explained in the text.}\label{3b2}
\end{figure}

The potentials in chiral dynamics can be split into an on-shell part which depends on the center of mass energy of  the interacting particles and an off-shell part proportional to $p^2 - m^2$ for each of the meson legs, in case of meson-meson interaction (where $p$ is the four vector of the off-shell particle and $m$ is its mass). In case of the meson-baryon interaction, the off-shell part of the potential behaves as $p^0 - k^0$, where $p^0 (k^0)$ is the energy corresponding to the off-shell (on-shell) momentum. Due to this behavior, the off-shell part of the potential cancels a propagator in the loops, giving rise effectively to a three body force, for example, the one shown in Fig. \ref{3b2} corresponding to the $t^1 g^{13} t^3$ term shown in Fig. \ref{3b}.

Similar effective three-body forces arise from other terms too. We shall now write the contributions for the first terms of all six $t^i g^{ij} t^j$ terms (Eq. (\ref{vgv})) including the off-shell parts of the $t$-matrices, taking the  $\pi^+ \pi^- n$ channel as an example and evaluate the total effect of these three-body forces.

\begin{figure}[ht!]
\centering
\includegraphics[width=0.4\textwidth]{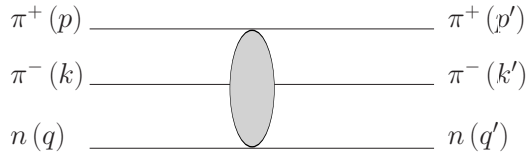}
\caption{\it Assigning four momenta to the $\pi^+\pi^-n$ system.}\label{pipiN}
\end{figure}

We label the initial (final) four-momentum of the $\pi^+$ as $p$ ($p^{\,\prime}$), that of the $\pi^-$ as $k$ $(k^{\,\prime})$ and that of the neutron as $q$ $(q^{\,\prime})$ as shown in Fig. \ref{pipiN}. We assign a four vector $k^{\prime\prime}$ to the intermediate states, see Fig. \ref{3b}

The potentials calculated from the chiral Lagrangians Eqs. (\ref{LMB}, \ref{LMM}) for the three possible two-body interactions are
\begin{equation}
V_{\pi^+ \pi^-\rightarrow \pi^+ \pi^-}=-\frac{1}{6f^2}\Big[3s_{\pi\pi}-\sum_i (p_i^2-m^2_i)\Big],\label{voff}
\end{equation}
\begin{equation}
V_{\pi^- n\rightarrow \pi^- n}=-\frac{1}{4f^2}(k_{\pi}^0 + {k_\pi^{\prime}}^0),
\end{equation}
\begin{equation}
V_{\pi^+ n\rightarrow \pi^+ n}=-V_{\pi^- n\rightarrow \pi^- n},
\end{equation}
where $s_{\pi\pi}$ is the invariant mass of the $\pi-\pi$ sub-system, $f$ is the pion decay constant and $k_{\pi}^0 \, ({k_\pi^{\prime}}^0)$ is the energy of the pion  before (after) the $\pi N$ interaction.

In this way, the contribution of the first term of Eq. (\ref{vgv}) for $i = 1$ and $j = 2$, which corresponds to the first diagram in Fig. \ref{alltgt}, is given by
\begin{equation}
T_a=-\frac{1}{16f^4}(k^0+{k^{\,\prime}}^0)\frac{m_n}{E_n}\frac{1}{{k^{\,\prime}}^0+{q^{\,\prime}}^0-k^0-E_n(\vec{p}^{\,\prime}+\vec{k}) + i\epsilon }(p^0+{p^{\,\prime}}^0)\equiv T^{on}_a.\label{ta}
\end{equation}
$m_n$ in Eq.(\ref{ta}) is the neutron mass and the superscript ``on'' on $T_a$ denotes that there is no off-shell dependence in the above equation.

For the diagram (b) of Fig. \ref{alltgt}
\begin{eqnarray}
T_b&=&\frac{1}{24f^4}\Big[2{k^{\,\prime}}^0+({k^{\prime\prime}}^0-{k^{\prime}}^0)\Big]\frac{1}{{k^{\prime\prime}}^2-m^2_\pi}\Big[3(p+k)^2-({k^{\prime\prime}}^2-m^2_\pi)\Big]\\
&\equiv& T_b^{on}+T_b^{off}
\end{eqnarray}
with
\begin{eqnarray}
T_b^{on}&=&\frac{1}{4f^4}{k^{\,\prime}}^0\frac{1}{(p+k-p^{\,\prime})^2-m^2_\pi}(p+k)^2 \label{tbon} \\
T_b^{off}&=&\frac{1}{24f^4}\Bigg[-{k^{\,\prime}}^0-p^0-k^0+{p^{\,\prime}}^0+3(p+k)^2\frac{{k^{\prime\prime}}^0-{k^{\,\prime}}^0}{{k^{\prime\prime}}^2-m^2_\pi}\Bigg] \label{tbof}
\end{eqnarray}
representing the on-shell and off-shell contributions to $T_b$. In Eq. (\ref{tbon}) and in the first term of Eq. (\ref{tbof}),  $k^{\prime\prime}$ has been replaced by $p + k - p^{\,\prime}$ using the energy-momentum conservation law from the initial state.
For the second term of Eq. (\ref{tbof}) we apply energy-momentum conservation from the final state
\begin{equation}
{k^{\prime\prime}}^2 = (k^{\,\prime} + q^\prime - q)^2. \label{deltaq}
\end{equation} 
Defining $\Delta q = q^{\,\prime} - q$, Eq. (\ref{deltaq}) becomes
\begin{equation}
{k^{\prime\prime}}^2 = m^2_\pi + (\Delta q)^2+2k^{\,\prime}\cdot\Delta q\nonumber
\end{equation}
and, hence
\begin{equation}
\frac{{k^{\prime\prime}}^0-{k^{\,\prime}}^0}{{k^{\prime\prime}}^2-m_\pi^2}=
\frac{(\Delta q)^0}{(\Delta q)^2+2k^{\,\prime}\cdot\Delta q}.\label{Delta}
\end{equation}
Therefore,
\begin{equation}
T_b^{off}=\frac{1}{24f^4}\Bigg[-{k^{\,\prime}}^0-p^0-k^0+{p^{\,\prime}}^0+3(p+k)^2\frac{(\Delta q)^0}{(\Delta q)^2+2k^{\,\prime}\cdot\Delta q}\Bigg].
\end{equation}
The contribution of the diagram (c) of Fig. \ref{alltgt} is
\begin{equation}
T_c=-\frac{1}{16f^4}(p^0+{p^{\,\prime}}^0)\frac{m_n}{E_n}\frac{1}{k^0+q^0-{k^{\,\prime}}^0-E_n(\vec{p}+\vec{k}^{\,\prime})}(k^0+{k^{\,\prime}}^0)\equiv T_c^{on}.
\end{equation}
Similarly,
\begin{eqnarray}
T_d&=&-\frac{1}{24f^4}(2{p^{\,\prime}}^0+{p^{\prime\prime}}^0-{p^{\prime}}^0)\frac{1}{{p^{\prime\prime}}^2-m^2_\pi}[3(p+k)^2-({p^{\prime\prime}}^2-m^2_\pi)]\\
&\equiv&T^{on}_d+T^{off}_d
\end{eqnarray}
with
\begin{eqnarray}
T_d^{on}&=&-\frac{1}{4f^4}{p^{\,\prime}}^0\frac{1}{(p+k-k^{\,\prime})^2-m^2_\pi}(p+k)^2\\
T_d^{off}&=&-\frac{1}{24f^4}\Bigg[-{p^{\,\prime}}^0-p^0-k^0+{k^{\,\prime}}^0+3(p+k)^2\frac{{p^{\prime\prime}}^0-{p^{\,\prime}}^0}{{p^{\prime\prime}}^2-m_\pi^2}\Bigg]
\end{eqnarray}
Analogous to Eq. (\ref{Delta}), we write
\begin{eqnarray}
\frac{{p^{\prime\prime}}^0-{p^{\,\prime}}^0}{{p^{\prime\prime}}^2-m_\pi^2}&=&\frac{(\Delta q)^0}{(\Delta q)^2+2p^{\,\prime}\cdot\Delta q}
\end{eqnarray}
which gives
\begin{equation}
T_d^{off}=-\frac{1}{24f^4}\Bigg[-{p^{\,\prime}}^0-p^0-k^0+{k^{\,\prime}}^0+3(p+k)^2\frac{(\Delta q)^0}{(\Delta q)^2+2p^{\,\prime}\cdot\Delta q}\Bigg].
\end{equation}
For the next diagram, we have
\begin{eqnarray}
T_e&=&\frac{1}{24f^4}[3(p^{\,\prime}+k^{\,\prime})^2-({k^{\prime\prime}}^2-m_\pi^2)]\frac{1}{{k^{\prime\prime}}^2-m_\pi^2}
(2k^0+{k^{\prime\prime}}^0-k^0)\\
&\equiv& T_e^{on}+T_e^{off}
\end{eqnarray}
where
\begin{eqnarray}
T_e^{on}&=&\frac{1}{4f^4}(p^{\,\prime}+k^{\,\prime})^2\frac{1}{(p^{\,\prime}+k^{\,\prime}-p)^2-m_\pi^2}k^0\\
T_e^{off}&=&\frac{1}{24f^4}\Bigg[-k^0-{p^{\,\prime}}^0-{k^{\,\prime}}^0+p^0+3(p^{\,\prime}+k^{\,\prime})^2\frac{{k^{\prime\prime}}^0-k^0}{{k^{\prime\prime}}^2-m_\pi^2}\Bigg]\nonumber
\end{eqnarray}
In this case $k^{\prime\prime}=k-\Delta q$ and therefore
\begin{equation}
\frac{{k^{\prime\prime}}^0-k^0}{{k^{\prime\prime}}^2-m_\pi^2}=-\frac{(\Delta q)^0}{(\Delta q)^2-2k\cdot\Delta q}
\end{equation}
leading to 
\begin{equation}
T_e^{off}=\frac{1}{24 f^4}\Bigg[-k^0-{p^{\,\prime}}^0-{k^{\,\prime}}^0+p^0-3(p^{\,\prime}+k^{\,\prime})^2\frac{(\Delta q)^0}{(\Delta q)^2-2k\cdot\Delta q}\Bigg].
\end{equation} 
For the last diagram of Fig. \ref{alltgt} we have
\begin{eqnarray}
T_f&=&-\frac{1}{24f^4}[3(p^{\,\prime}+k^{\,\prime})^2-({p^{\prime\prime}}^2-m_\pi^2)]\frac{1}{{p^{\prime\prime}}^2-m_\pi^2}
(2p^0+{p^{\prime\prime}}^0-p^0)\\
&\equiv& T_f^{on}+T_f^{off}
\end{eqnarray}
where
\begin{eqnarray}
T_f^{on}&=&-\frac{1}{4f^4}(p^{\,\prime}+k^{\,\prime})^2\frac{1}{(p^{\,\prime}+k^{\,\prime}-k)^2-m_\pi^2}p^0\\
T_f^{off}&=&-\frac{1}{24f^4}\Bigg[-p^0-{p^{\,\prime}}^0-{k^{\,\prime}}^0+k^0+3(p^{\,\prime}+k^{\,\prime})^2\frac{{p^{\prime\prime}}^0-p^0}{{p^{\prime\prime}}^2-m_\pi^2}\Bigg]
\end{eqnarray}
Following the same method
\begin{equation}
\frac{{p^{\prime\prime}}^0-p^0}{{p^{\prime\prime}}^2-m_\pi^2}=-\frac{(\Delta q)^0}{(\Delta q)^2-2p\cdot\Delta q}.
\end{equation}
then
\begin{equation}
T_f^{off}=-\frac{1}{24f^4}\Bigg[-p^0-{p^{\,\prime}}^0-{k^{\,\prime}}^0+k^0-3(p^{\,\prime}+k^{\,\prime})^2\frac{(\Delta q)^0}{(\Delta q)^2-2p\cdot\Delta q}\Bigg].
\end{equation}

\begin{figure}
\centering
\includegraphics[scale=0.7]{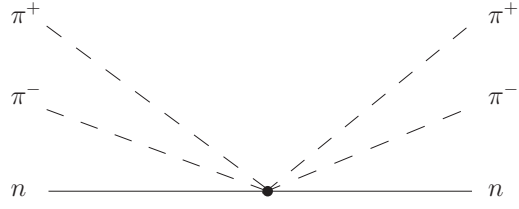}
\caption{\it Source of three-body force from the chiral Lagrangians.}\label{contact}
\end{figure}

On the other hand, genuine three-body forces also originate directly from the chiral Lagrangian , where we can find a contact term as the one shown in Fig. (\ref{contact}) \cite{felipe}. 

At lowest order in momentum, which we consider in our study, the interaction Lagrangian between mesons and baryon is given by
\begin{equation}
\mathcal{L}=i\langle \bar{B}\gamma^\mu[\Gamma_\mu,B]\rangle\label{Lag}
\end{equation}
where 
\begin{equation}
\Gamma_\mu=\frac{1}{2}(u^\dagger\partial_\mu u+u\partial_\mu u^\dagger), \quad u^2=e^{i \sqrt{2} \Phi/f }
\end{equation}
and $\phi$, $B$ are same as those in Eqs. (\ref{eqb}, \ref{phiBM}).
If we expand $\Gamma_\mu$ up to the terms which contain four meson fields, we get
\begin{equation}
\Gamma_\mu=\frac{1}{32f^4}\Bigg[\frac{1}{3}\partial_\mu\Phi\Phi^3-\Phi\partial_\mu\Phi\Phi^2+\Phi^2\partial_\mu\Phi\Phi-\frac{1}{3}\Phi^3\partial_\mu\Phi\Bigg]
\end{equation}
For the case under consideration, i.e., $\pi^+\pi^- n$, the Eq. (\ref{Lag}) becomes
\begin{equation}
\mathcal{L}=\frac{i}{32 f^4}\bar{n}\Bigg[\frac{1}{3}\sla \partial \pi^-\pi^+\pi^-\pi^+ -\pi^-\sla \partial\pi^+\pi^-\pi^+ +\pi^-\pi^+\sla \partial\pi^-\pi^+-\frac{1}{3}\pi^-\pi^+\pi^-\sla \partial\pi^+\Bigg]n
\end{equation}
In this way, the contribution of the diagram in Fig. \ref{contact} is
\begin{equation} 
T_{3b}=\frac{1}{24f^4}\bar{u}_r(\vec{q}^{\,\prime})(2\sla p-2\sla k^{\,\prime}-2\sla k+2\sla p^{\,\prime})u_r(\vec{q})\label{T3b}
\end{equation}
We are interested in the low energy region, thus, only the $\gamma^0$ component of Eq. (\ref{T3b}) is relevant, then
\begin{equation} 
T_{3b}=\frac{1}{24f^4}(2p^0-2{k^{\,\prime}}^0-2k^0+2{p^{\,\prime}}^0)
\end{equation}
Adding this to the off-shell contributions from the Faddeev equations at second order in $t$-matrices, we get
\begin{eqnarray}
\sum\limits_{i=1}^6 T_i^{off}+T_{3b}&=&\frac{1}{24f^4}\Bigg[-4k^0+4{p^{\,\prime}}^0-4{k^{\,\prime}}^0+4p^0\nonumber\\
&&+3(p+k)^2(\Delta q)^0\Bigg\{\frac{1}{(\Delta q)^2+2k^{\,\prime}\cdot\Delta q}-\frac{1}{(\Delta q)^2+2p^{\,\prime}\cdot\Delta q}\Bigg\}\nonumber\\
&&+3(p^{\,\prime}+k^{\,\prime})^2(\Delta q)^0\Bigg\{\frac{1}{(\Delta q)^2-2p\cdot\Delta q}-\frac{1}{(\Delta q)^2-2k\cdot\Delta q}\Bigg\}\Bigg]\label{sum}
\end{eqnarray}
If we consider small momentum transfer for the baryon, i.e., $\Delta \vec{q}<< 1$, Eq. (\ref{sum}) can be expressed as
\begin{eqnarray}
\sum\limits_{i=1}^6 T_i^{off}+T_{3b}&=&\frac{1}{24f^4}\Bigg[-4k^0+4{p^{\,\prime}}^0-4{k^{\,\prime}}^0+4p^0+3(p+k)^2\Bigg\{\frac{1}{(\Delta q)^0+2{k^{\,\prime}}^0}-\frac{1}{(\Delta q)^0+2{p^{\,\prime}}^0}\Bigg\}\nonumber\\
&&+3(p^{\,\prime}+k^{\,\prime})^2\Bigg\{\frac{1}{(\Delta q)^0-2p^0}-\frac{1}{(\Delta q)^0-2k^0}\Bigg\}\Bigg]
\end{eqnarray}
And there is a cancellation of the terms in the $SU(2)$ limit, assuming equal average energies for the pion. Furthermore, if the propagators in the Eq. (\ref{sum}) are projected over s-wave, as we do in our study, the curly brackets become
\begin{eqnarray}
\Biggr \{ \frac{1}{2\mid \vec{k}^{\,\prime} \mid \, \mid \vec{\Delta q} \mid} ln \Biggr( \frac{(\Delta q)^2 + 2 {k^{\,\prime}}^0 (\Delta q)^0 + 2\mid \vec{k}^{\,\prime} \mid \, \mid \vec{\Delta q} \mid}{(\Delta q)^2 + 2 {k^{\,\prime}}^0(\Delta q)^0 - 2\mid \vec{k}^{\,\prime} \mid \, \mid \vec{\Delta q}\mid} \Biggr)   \nonumber \\
- \, \frac{1}{2\mid \vec{p}^{\,\prime} \mid \, \mid \vec{\Delta q} \mid} ln \Biggr( \frac{(\Delta q)^2 + 2 {p^{\,\prime}}^0 (\Delta q)^0 + 2\mid \vec{p}^{\,\prime} \mid \, \mid \vec{\Delta q} \mid} {(\Delta q)^2 + 2 {p^{\,\prime}}^0 (\Delta q)^0 - 2\mid \vec{p}^{\,\prime} \mid \, \mid \vec{\Delta q}\mid}\Biggr) \Biggr \} \nonumber
\end{eqnarray}
and 
\begin{eqnarray}
\Biggr \{ \frac{1}{2\mid \vec{p} \mid \, \mid \vec{\Delta q} \mid} ln \Biggr( \frac{(\Delta q)^2 - 2 p^0 (\Delta q)^0 + 2\mid \vec{p} \mid \, \mid \vec{\Delta q} \mid}{(\Delta q)^2 - 2 p^0 (\Delta q)^0 - 2\mid \vec{p} \mid \, \mid \vec{\Delta q}\mid} \Biggr)  \nonumber \\
- \, \frac{1}{2\mid \vec{k} \mid \, \mid \vec{\Delta q} \mid} ln \Biggr( \frac{(\Delta q)^2 - 2 k^0 (\Delta q)^0 + 2\mid \vec{k} \mid \, \mid \vec{\Delta q} \mid} {(\Delta q)^2 -2 k^0 (\Delta q)^0 - 2\mid \vec{k} \mid \, \mid \vec{\Delta q}\mid}\Biggr) \Biggr \} \nonumber
\end{eqnarray}
respectively, and the cancellation is exact.

With the cancellation of these basic diagrams proved, the addition of an extra interaction, $v^i$, to the set of these cancelling terms will still give a vanishing contribution. For the on-shell part of the basic diagrams that we have studied, further iteration of the potential between two particles, leading to the two-body t-matrix, are done in such a way that the off-shell part of the new $v^i$ interactions is reabsorbed in constants of the on-shell potential, as is done in the construction of the two body $t$-matrices \cite{angels, npa}. The conclusion is that at the end we should only use the on-shell $t$-matrices, ignoring the off-shell effects and genuine three body forces simultaneously.


\begin{thebibliography}{99}
 
\bibitem{pdg} W.-M. Yao {\it et al.}, J. Phys. G \textbf{33} (2006) 1 .

\bibitem{isgur1}
  N.~Isgur and G.~Karl,
  Phys.\ Rev.\  D {\bf 18} (1978) 4187.

\bibitem{isgur2}
  N.~Isgur and G.~Karl,
  Phys.\ Rev.\  D {\bf 19} (1979) 2653
  [Erratum-ibid.\  D {\bf 23} (1981) 817].

\bibitem{glozman}
  L.~Y.~Glozman and D.~O.~Riska,
  Phys.\ Rept.\  {\bf 268} (1996) 263
  [arXiv:hep-ph/9505422].


\bibitem{arndt1}
R. A. Arndt, W. J. Briscoe, I. I. Strakovsky, and R. L. Workman, 
Phys. Rev. C {\bf 69}  (2004) 035213 .


\bibitem{arndt2}
R. A. Arndt, W. J. Briscoe, I. I. Strakovsky, and R. L. Workman, 
Phys. Rev. C {\bf 74}  (2006) 045205.

\bibitem{cut}
R. E. Cutkosky, C. P. Forsyth, R. E. Hendrick, and R. L. Kelly, Phys. Rev. D 20 (1979) 2839. 

\bibitem{manley}
D. M. Manley and E. M. Saleski, Phys. Rev. D {\bf 45} (1992) 4002.

\bibitem{chiang}
Wen-Tai Chiang, B. Saghai, F. Tabakin, and T. S.-H. Lee, Phys. Rev. C {\bf 69} (2004) 065208 .

\bibitem{batinic}
M. Batinic, I. Slaus, A. Svarc, and B. M. K. Nefkens, Phys. Rev. C {\bf 51} (1995) 2310;
M. Batinic, I. Dadic, I. Slaus, A. Svarc, B. M. K. Nefkens, and T. S.-H. Lee, Phys. Scr. 58 (1998) 15.


\bibitem{ceci}
  S.~Ceci, A.~Svarc and B.~Zauner,
  Phys.\ Rev.\ Lett.\  {\bf 97} (2006) 062002
  [arXiv:hep-ph/0603144].


\bibitem{nakano}
  T.~Nakano {\it et al.}  [LEPS Collaboration],
  Phys.\ Rev.\ Lett.\  {\bf 91} (2003) 012002
  [arXiv:hep-ex/0301020].

\bibitem{diakonov}
  D.~Diakonov, V.~Petrov and M.~V.~Polyakov,
  Z.\ Phys.\  A {\bf 359} (1997) 305
  [arXiv:hep-ph/9703373].


\bibitem{jaffe}
R. Jaffe and F. Wilczek, Phys. Rev. Lett. {\bf 91} (2003) 232003.


\bibitem{ceci2}
  S.~Ceci, A.~Svarc and B.~Zauner,
  Few Body Syst.\  {\bf 39} (2006) 27
  [arXiv:hep-ph/0512337].


\bibitem{arndt3}
  R.~A.~Arndt, Y.~I.~Azimov, M.~V.~Polyakov, I.~I.~Strakovsky and R.~L.~Workman,
  Phys.\ Rev.\  C {\bf 69} (2004) 035208
  [arXiv:nucl-th/0312126].

\bibitem{arndt4}
  I.~I.~Strakovsky, R.~A.~Arndt, Y.~I.~Azimov, M.~V.~Polyakov and R.~L.~Workman,
  J.\ Phys.\ Conf.\ Ser.\  {\bf 9} (2005) 218
  [arXiv:hep-ph/0501114].

\bibitem{hicks}
  K.~H.~Hicks,
  Prog.\ Part.\ Nucl.\ Phys.\  {\bf 55} (2005) 647
  [arXiv:hep-ex/0504027].


\bibitem{q3g} 
Z. P. Li, Phys. Rev. D {\bf 44} (1991) 2841;
Z. P. Li, V. Burkert, and Zh. Li, Phys. Rev. D {\bf 46} (1992) 70.

\bibitem{cano}
F. Cano, P. Gonz´alez, S. Noguera, and B. Desplanques,
Nucl. Phys. A {\bf 603} (1996) 257;
F. Cano and P. Gonz´alez, Phys. Lett. B {\bf 431} (1998) 270.


\bibitem{juelich}
  O.~Krehl, C.~Hanhart, S.~Krewald and J.~Speth,
  Phys.\ Rev.\  C {\bf 62} (2000) 025207
  [arXiv:nucl-th/9911080].


\bibitem{dillig}
M. Dillig and J. Schott, Phys. Rev. C {\bf 75} (2007) 067001.


\bibitem{hosaka}
  A.~Hosaka, T.~Hyodo, F.~J.~Llanes-Estrada, E.~Oset, J.~R.~Pelaez and M.~J.~Vicente Vacas,
  Phys.\ Rev.\  C {\bf 71} (2005) 045205
  [arXiv:hep-ph/0411311].


\bibitem{hyodo}
  T.~Hyodo, A.~Hosaka, E.~Oset, A.~Ramos and M.~J.~Vicente Vacas,
  Phys.\ Rev.\  C {\bf 68} (2003) 065203
  [arXiv:nucl-th/0307005].



\bibitem{mko1}
  A.~Martinez Torres, K.~P.~Khemchandani and E.~Oset,
 Phys. Rev. C {\bf 77} (2008) 042203, arXiv:0706.2330 [nucl-th].
  
\bibitem{mko2}
  A.~Martinez Torres, K.~P.~Khemchandani and E.~Oset,
 arXiv:0706.2330 [nucl-th]; 
{\it ibid} Eur. Phys. J. A {\bf 35} (2008) 295.

\bibitem{mko3}
  A.~Martinez Torres, K.~P.~Khemchandani, L.~S.~Geng, M.~Napsuciale and E.~Oset,
  arXiv:0801.3635 [nucl-th].

\bibitem{BABAR1}
  B.~Aubert {\it et al.}  [BABAR Collaboration],
  Phys.\ Rev.\  D {\bf 74} (2006) 091103
  [arXiv:hep-ex/0610018].
  
\bibitem{BABAR2}
  B.~Aubert {\it et al.}  [BABAR Collaboration],
  Phys.\ Rev.\  D {\bf 76} (2007) 012008
  [arXiv:0704.0630 [hep-ex]].


\bibitem{BES}
M. Ablikim   {\it  et al.}  [BES Collaboration],
  arXiv:0712.1143 [hep-ex]. X.Y. Shen et al, International Workshop on Heavy
   Quarkonium 2007, 17-20 October 2007, DESY Hamburg,
http://www.desy.de/qwg07/agenda.php

\bibitem{weinberg}
S. Weinberg, Phys. Rev. {\bf 166} (1968), (1968) 1568.


\bibitem{meissner}
V.~Bernard, N.~Kaiser and U.~G.~Meissner,
  Int.\ J.\ Mod.\ Phys.\  E {\bf 4} (1995) 193
  [arXiv:hep-ph/9501384].

\bibitem{ecker}
  G.~Ecker,
  Prog.\ Part.\ Nucl.\ Phys.\  {\bf 35} (1995) 1
  [arXiv:hep-ph/9501357].


\bibitem{angels}
  E.~Oset and A.~Ramos,
  Nucl.\ Phys.\  A {\bf 635} (1998) 99.

\bibitem{Inoue}
  T.~Inoue, E.~Oset and M.~J.~Vicente Vacas,
  Phys.\ Rev.\  C {\bf 65} (2002) 035204
  [arXiv:hep-ph/0110333].


\bibitem{npa}
  J.~A.~Oller and E.~Oset,
  Nucl.\ Phys.\  A {\bf 620} (1997) 438
  [Erratum-ibid.\  A {\bf 652} (1999) 407]
  [arXiv:hep-ph/9702314].

\bibitem{oller}
  J.~A.~Oller, E.~Oset and J.~R.~Pelaez,
  Phys.\ Rev.\  D {\bf 59} (1999) 074001
  [Erratum-ibid.\  D {\bf 60} (1999\ ERRAT,D75,099903.2007) 099906]
  [arXiv:hep-ph/9804209].


\bibitem{mandlshaw}
F. Mandl and G. Shaw, {\it Quantum Field Theory}  Wiley-Interscience Publication.

\bibitem{Fernandez}
  P.~Fernandez de Cordoba, Yu.~Ratis, E.~Oset, J.~Nieves, M.~J.~Vicente-Vacas,
  B.~Lopez-Alvaredo and F.~Gareev,
  Nucl.\ Phys.\  A {\bf 586} (1995) 586 .

\bibitem{eli}
  E.~Hernandez, E.~Oset and M.~J.~Vicente Vacas,
  Phys.\ Rev.\  C {\bf 66} (2002) 065201
  [arXiv:nucl-th/0209009].


\bibitem{felipe}
  F.~J.~Llanes-Estrada, E.~Oset and V.~Mateu,
  Phys.\ Rev.\  C {\bf 69} (2004) 055203
  [arXiv:nucl-th/0311020].
\end{thebibliography}
\end{document}